\documentclass[runningheads]{llncs}
\usepackage[T1]{fontenc}
\usepackage{graphicx}
\usepackage{braket}
\usepackage{graphics}
\usepackage{amssymb}
\usepackage{amsmath, framed}
\usepackage{braket}
\usepackage[ruled,vlined,linesnumbered]{algorithm2e}
\SetKwComment{tcp}{\small$\triangleright$\,}{}
\usepackage{epsfig}
\usepackage{amsmath}
\usepackage{amsfonts}
\usepackage{amssymb}
\usepackage{enumerate}
\usepackage{booktabs}
\usepackage{multirow}
\usepackage{tabularx}
\usepackage{array}
\usepackage{enumitem}

\usepackage{circuitikz}

\usepackage{color,soul}

\usepackage{pdflscape}
\usepackage{longtable}
\usepackage{graphicx}
\usepackage{fancyhdr} 

\newcommand{\Tr}{\text{Tr}}
\usepackage{amsmath, amssymb}

\newcommand{\Bcal}{\mathcal{B}}

\newcommand{\Hilb}{\mathcal{H}}
\usepackage{booktabs,multirow,array}
\usepackage{physics}           
\usepackage{mathtools}         
\usepackage{tikz-cd}           

\usepackage{ifpdf}
\usepackage{qcircuit}
\usepackage{pifont}
\usepackage{graphicx}
\usepackage{braket}
\usepackage{pifont}
\usepackage{array}
\usepackage{longtable}
\usepackage{bm}
\usepackage{hyperref}
\usepackage{color}
\usepackage{makeidx}
\usepackage{cite}
\usepackage{mathtools}
\usepackage{bbm}
\usepackage{wrapfig}

\usepackage{amsmath, amssymb}    
\usepackage{tabularx}            
\usepackage{booktabs}            

\usepackage{tikz}                
\usetikzlibrary{arrows.meta}     
\usetikzlibrary{positioning}     

\usepackage{xcolor}              
\begin{document}
\title{Hamiltonian Formalism for Comparing Quantum and Classical Intelligence}
\titlerunning{Hamiltonian Intelligence}
%
\author{Elija Perrier\inst{1}\orcidID{0000-0002-6052-6798} }
\institute{Centre for Quantum Software \& Information, UTS, Sydney \email{elija.perrier@gmail.com}}

%

%

%
\maketitle              
\begin{abstract}
The prospect of AGI instantiated on quantum substrates motivates the development of mathematical frameworks that enable direct comparison of their operation in classical and quantum environments. To this end, we introduce a Hamiltonian formalism for describing classical and quantum AGI tasks as a means of contrasting their interaction with the environment. We propose a decomposition of AGI dynamics into Hamiltonian generators for core functions such as induction, reasoning, recursion, learning, measurement, and memory. This formalism aims to contribute to the development of a precise mathematical language for how quantum and classical agents differ via environmental interaction.\footnote{This is the author’s version accepted at AGI-25 (camera-ready length limit = 10 pages plus references/appendices).
Follow-on work detailing bounds and limitations is in preparation; comments/criticisms welcome.}

\end{abstract}
\section{Introduction}
Interest in the synthesis of classical artificial general intelligence \cite{goertzel2014,hutter2004universal} with emerging quantum information processing (QIP) \cite{NielsenChuang2010,aaronson2013quantum,watrous_theory_2018,preskill_quantum_2021} technologies has given rise to questions regarding how the underlying physical substrate upon which intelligent systems are constructed influences their nature and capabilities. Most AGI theories are classical: implicitly assuming a computational and informational model grounded in classical physics \cite{RussellNorvig2020,bennettmaruyama2022b,sole2019,goertzel2021,goertzel2023,mcmillen2024}. Yet quantum mechanics offers a profoundly different ontology \cite{ozkural2012like,bell_speakable_2004,sakurai2020modern,feynman_simulating_1982,manin2007mathematics} due to phenomena such as superposition, entanglement, non-locality, contextuality \cite{kochenspecker1967} and no-cloning \cite{wootters_single_1982}. Hamiltonian mechanics \cite{goldstein_classical_2002} offers a powerful and unifying language to describe the dynamics of both classical and quantum systems. In this work, we use Hamiltonian dynamics to model AGI in classical AGI (CAGI) quantum AGI (QAGI) settings. We demonstrate how key AGI functionalities can be associated with specific Hamiltonian generators. The algebraic properties of these generators (e.g., their commutation relations) affect the capabilities of each respective AGI, influencing their capacity for information processing, logical reasoning, learning, and interaction. By analysing these structures, we aim to contribute to the development of a mathematically rigorous theory of quantum agency.
\\
\\
\textbf{Background \& Related Work}. 
Classical systems evolve on a symplectic phase space manifold $M$ \cite{hall_quantum_2013}. Observables are represented by smooth functions $f \in C^\infty(M)$, while their dynamics are represented variationally via Hamiltonian dynamics using Poisson brackets $\dot f = \{ f,H\}$. Quantum systems, by contrast, are described by states in a Hilbert space $\Hilb$, with observables represented by self-adjoint operators. Their dynamics are governed by the Schrödinger equation, and are inherently tied to the non-commutative algebra of these operators. This non-commutativity is accompanied by quantum phenomena such as superposition, entanglement, measurement stochasticity, and contextuality \cite{bell_speakable_2004,kochenspecker1967}.  Adopting an approach that synthesises concepts from quantum information theory \cite{watrous_theory_2018}, quantum circuit formalism \cite{chiribella2008quantum} and geometry \cite{chruscinski_geometric_2004,knapp_lie_1996,helgason_differential_1979,frankel_geometry_2011,perrier2024quantum}, we conceptualise an agent's cognitive and interactive processes as arising from a set of fundamental Hamiltonian generators. This approach allows for, in certain circumstances, a direct comparison of agent/environment interactions where agents and/or environments may be classical and quantum. By introducing a variational-based approach, we can in principle analyse structures that govern their respective dynamics and differences.
%
\\
\\
\textbf{Classical and Quantum Information Processing}.
To compare CAGI and QAGI, we require a means of formulating them both within a common theoretical framework. However, doing so is not simple. Quantum mechanics and classical mechanics, while sharing considerable overlap, are fundamentally different in important ways that affect their comparison. For agents, this problematises common classical assumptions regarding identifiability, certainty of state descriptions and the distinguishability of an agent from its environment. To formulate CAGI and QAGI in a way that shines light on their differences, we frame both systems in the paradigm of quantum information theory \cite{watrous_theory_2018} (QIP). In this formulation, agents and environments are described via informational \textit{registers} $\texttt{X}$ (e.g. bits) comprising information drawn from a classical alphabet $\Sigma$. The states of registers may be either classical or quantum states. We define a Hilbert space $\mathcal X=\mathbb \mathbb{C}^{|\Sigma|}$ with computational basis 
$\{|s\rangle\}_{s\in\Sigma}$. A \textit{quantum} state of a register \texttt{X} (associated with space $\mathcal{X}$) is a density operator $\rho \in \mathcal{D}(\mathcal{X})$, i.e., a positive semi-definite operator with $\Tr(\rho)=1$. \textit{Classical} states are precisely those density operators
that are diagonal in the distinguished computational basis
$\{\ket{s}\}_{s\in\Sigma}$ of $\mathcal X$
(i.e.\ $\rho = \sum_{s\in\Sigma} p_s\ket{s}\!\bra{s}$ with
$p_s\ge 0,\;\sum_s p_s = 1$). Interactions to (and changes of) states occur via \textit{channels} which are superoperators. They define how quantum and classical states interact. Classical-to-classical channels (CTC) preserve classical states, classical-to-quantum (CTQ) channels encode classical information in quantum states, quantum-to-classical (QTC) channels extract classical information from quantum states, decohering them in the process; quantum-to-quantum (QTQ) channels form coherent (e.g. unitary) transformations between quantum registers. In this framing, both agents and environments are registers which may be CAGI (classical state sets) or QAGI (quantum state sets). They may interact in ways that are coherent (quantum-preserving) or classical (via CTC or QTC maps). These channels act on the algebra of observables: CTC preserves commutative subalgebras, QTQ preserves the full non-commutative structure, while CTQ/QTC mediate between them. A diagrammatic illustration is set out Figures \ref{fig:CAGI} and \ref{fig:QAGI} in the Appendix.
\\
\\
\textbf{Classical AGI Hamiltonians}.
Using a classical mechanical paradigm, CAGI can be conceptualised as a dynamical system evolving in a high-dimensional phase space $M = T^*\mathcal{C}$, the cotangent bundle of its configuration space $\mathcal{C}$. The state of the AGI at any time is given by a point $(\mathbf{q}, \mathbf{p}) \in M$, where $\mathbf{q} = (q_1, \dots, q_n)$ are generalized coordinates representing, for instance, memory contents, internal model parameters, or sensor readings, and $\mathbf{p} = (p_1, \dots, p_n)$ are their conjugate momenta, representing rates of change or dynamic aspects. Observables may be smooth real-valued functions $f(\mathbf{q}, \mathbf{p})$ on this phase space, but they are not always. The dynamics of AGI we model as being governed by a total Hamiltonian $H_C(\mathbf{q}, \mathbf{p})$, a function representing the AGI's total energy or a cost function to be optimized. Evolution is described by Hamilton's equations which can succinctly be represented using Poisson-bracket formalism $\dot{f} = \{f, H_C\}_{PB}$.  If $\{f, g\}_{PB} = 0$ (Poisson bracket), the observables $f$ and $g$ are said to commute, implying they can, in principle, be simultaneously determined with arbitrary precision. Classical logic and computation often implicitly rely on this property: the truth value of one proposition or the state of one register does not inherently interfere with another, distinct one unless explicitly coupled by $H_C$. For a classical AGI, $H_C$ we model Hamiltonians as decomposable: $H_C = \sum_k H_{C,k}$, where each $H_{C,k}$ represents a functional aspect like learning (e.g., gradient descent dynamics \cite{sunehag2012optimistic}), reasoning (e.g., energy function of a Hopfield network or constraint satisfaction), or interaction. The commutativity of these underlying processes, or the variables they act upon, defines the classical computational semantics.
\\
\\
\textbf{Quantum AGI Hamiltonians}. When transitioning to a quantum substrate, the AGI's state is described by a vector $\ket{\psi}$ in a Hilbert space $\Hilb$ (or a density operator $\rho$ acting on $\Hilb$). Observables are represented by self-adjoint operators $A$ acting on $\Hilb$. The dynamics are governed by the Schr\"odinger equation $i\hbar \frac{\dd}{\dd t}\rho(t) = [H_Q, \rho(t)]$ where $H_Q$ is the quantum Hamiltonian operator and $[A,B] = AB - BA$ is the commutator. The key algebraic difference from the classical case lies in the non-commutativity of operators when $[A,B] \neq 0$, giving rise to consequences explored below. For a quantum AGI, the total Hamiltonian $H_Q = \sum_k H_{Q,k}$ would similarly consist of generators for different AGI functions. However, these $H_{Q,k}$ are now operators, and their mutual commutation relations, as well as their commutation with other relevant observables, dictate the AGI's behavior. For example, if a learning operator $H_{Q,learn}$ does not commute with a sensing operator $H_{Q,sens}$ representing environmental perception, then the act of learning can be disturbed by observation, and vice-versa, in a way that has no classical parallel. This non-commutative structure underpins quantum phenomena like entanglement and contextuality. More background is set out in the Appendix. 
%
\begin{table}[ht!]
\centering\small
\caption{Example phase–space coordinates interpreted as measurable AGI features.}
\begin{tabular}{@{}>{\raggedright}p{2cm} p{3.3cm} p{3.3cm} p{3.1cm}@{}}
\toprule
Coordinate & CAGI & QAGI & Observable property \\
\midrule
$q_i$ (configuration) &
Serialized memory block, weight checkpoint, sensor pixel &
Expectation $\langle O_i\rangle_\rho$ of a Hermitian register operator $O_i$ via POVM tomography &
Used to infer current cognitive state (beliefs, activations, sensor snapshot) \\
\hline

$p_i = \dot q_i m$ (momentum) &
Finite-difference or profiler-level derivative of $q_i$ where $m$ is a weighting (e.g. learning rate) hyperparameter &
Symmetric logarithmic derivative $L_i$ &
Helps learn speed, attention switch-rate, reaction latency \\
\hline

$H(\mathbf q,\mathbf p)$ (Hamiltonian cost) &
Run-time energy proxy (FLOP budget, Joule counter, cross-entropy loss) as a function of generalised coordinates &
Generator of flow $\mathcal L(\rho) = i/\hbar[H,\rho] + ...$ reconstructed from process-tomography&
Indicator of resource consumption or cost of cognitive updates \\
\hline

$\Omega=\sum dq\wedge dp$ &
Liouville-volume tracker determinant of Jacobian of map $(q,p) \mapsto (q',p')$ &
Curvature measures &
Can measure information conservation / rates of change \\
\hline

Entropy $S(\mathbf q)$ &
Shannon entropy of probability over hidden-state ensemble &
von Neumann entropy $S(\rho) = -\Tr \rho \log \rho$ &
Uncertainty of the agent’s belief or internal noise level\\
\hline

Fisher / Bures metric  &
Fisher–Rao on parameter manifold (natural-gradient log) on parameters &
Bures metric via quantum Fisher information &
Sensitivity / curvature $\to$ expected generalisation \\
\bottomrule
\end{tabular}
\label{tab:ops-phase}
\end{table}

\vspace{-2em}
\section{Generator Decomposition Analysis}
\label{sec:generator_decomposition}
We now decompose the total AGI Hamiltonian in order to compare CAGI and QAGI acting in various environments. For each $H_G$, we contrast $H_G^C$ (classical) and $H_G^Q$ (quantum). Boldface denotes a vector (multi-degree-of-freedom object), plain italic a single coordinate (unless otherwise indicated). Classical phase space is $T^*\mathcal{C}$ with $(\mathbf{q},\mathbf{p})$; quantum states $\rho$ are on $\Hilb_A \otimes \Hilb_E$. Pauli operators on logical qubits are $X_k, Y_k, Z_k$. We set $\hbar=1$. It is useful to build intuition at this stage for exactly what the observables in the classical case may be. Table \ref{tab:ops-phase} offers a (non-exhaustive) prospective set of generalised coordinates and conjugate momenta in terms of \emph{instrumentable observables} that may be used in order to construct phase--space coordinates
$(\mathbf q,\mathbf p)$. Each
row specifies (i) how a CAGI agent logs or senses the variable,
(ii) the quantum measures the observable, and (iii) the relevant AGI property of interest that can be inferred.\\
\\
\textbf{Induction}. Induction in our framework represents the process by which an agent updates its internal model based on observed data. In the classical case, this corresponds to parameter optimization via gradient descent on prediction error, which we cast in Hamiltonian form by treating the loss function as a potential energy and introducing momentum terms for parameter dynamics. In CAGI, induction corresponds to minimising error on the statistical manifold parametrized by $f_{\boldsymbol\theta}$. To model this process, we use $H_{\mathrm{ind}}$ as the generator measuring information in data.  In QAGI it becomes the
relative‐entropy distance on state space. Information-geometrically the classical term measures Fisher length,
while the quantum term measures Bures length. The two coincide whenever
$\rho_D$ and $\rho_\theta$ commute.

\paragraph{Classical form ($H_{\mathrm{ind}}^C$).} Given data $\mathcal{D}=\{(\mathbf{s}_i,\mathbf{r}_i)\}_{i=1}^{N}$, model $f_{\boldsymbol{\theta}}$, weights $w_i$:
\begin{equation}
  H_{\mathrm{ind}}^{C} = \sum_{i=1}^{N} \frac{w_i}{2}\, \bigl\| f_{\boldsymbol{\theta}}(\mathbf{s}_i)-\mathbf{r}_i \bigr\|_2^{2} + \sum_{\ell=1}^{|\boldsymbol{\theta}|} \frac{p_{\theta_\ell}^{2}}{2m_\ell}.
  \label{eq:H-ind-class-paper3}
\end{equation}
where $f_{\boldsymbol\theta}$ is a parametric predictor with weights $\boldsymbol\theta$ and sample-weights $w_i$;  
$p_{\theta_\ell}$ is the momentum conjugate to $\theta_\ell$ with weight $m_\ell$ (measured by finite differences in a log of $\theta_\ell(t)$, akin to a momentum term). This can model gradient descent dynamics for learning parameters in AIXI-like agents \cite{hutter2004universal,veness2012ensemble} or other inductive systems \cite{solomonoff1964,potapov2014making}. 

\paragraph{Quantum form ($H_{\mathrm{ind}}^Q$).} The quantum analogue replaces classical prediction error with quantum relative entropy between the empirical data state $\rho_{\!\mathcal{D}}$ and $\rho_{\boldsymbol{\theta}}$ the agent's predictive state, capturing how quantum learning must respect fundamental trade-offs imposed by non-commuting observables. Using relative entropy $S(\rho_1\|\rho_2) = \Tr[\rho_1(\ln\rho_1 - \ln\rho_2)]$:
\begin{equation}
  H_{\mathrm{ind}}^{Q} = k_{\mathrm{B}}T\, S\!\bigl( \rho_{\!\mathcal{D}}\|\rho_{\boldsymbol{\theta}} \bigr).
  \label{eq:H-ind-quant-paper3}
\end{equation}
where $k_{\mathrm B}T$ rescales quantum relative entropy $S(\rho_1\Vert\rho_2)=\Tr[\rho_1(\log\rho_1-\log\rho_2)]$ into energetic units. Generally, $[H_{\mathrm{ind}}^{Q},\rho_{\!\mathcal{D}}]\neq0$ if $\rho_{\!\mathcal{D}}$ and $\rho_{\boldsymbol{\theta}}$ don't commute. This implies that the act of learning (reducing relative entropy) can disturb the evidence state $\rho_{\!\mathcal{D}}$. This contrasts with classical Solomonoff induction where the data sequence is fixed \cite{hutter2024introduction}. Note that $H_{\mathrm{ind}}^{C}$ serves as a variational principle for parameter evolution, not energy conservation. Energy in this case reflects a computational resource (cost) that the learning dynamics minimise through dissipative gradient flow, rather than a conserved quantity.
%
%
\\
\\
\textbf{Reasoning---Logical Consistency ($H_{\mathrm{reas}}$)}. Reasoning can be modelled via $H_{\mathrm{reas}}$, a penalty term that
encodes consistency with logical rules. Logical clauses are encoded as energy penalties where violations of logical constraints increase the system's energy, naturally driving the agent toward logically consistent states. The ground subspace of $H_{\text{reas}}$ corresponds to assignments satisfying all constraints. In classical systems, logical propositions can be evaluated independently and combined without interference, corresponding to the commutative nature of Boolean operations. In quantum settings, non-commuting projectors in quantum logic or semantics mean the truth of one clause depends on which other clauses are measured first due to
contextuality. Denote $\mathcal C$ as the agent’s configuration manifold: the set of all instantaneous values of its state variables (weights, memory cells, sensor registers) with $p \in T^*\mathcal C$ its cotangent bundle. $\varphi_\alpha$ is a Boolean predicate evaluating to 1 when the clause is satisfied in the current classical state. $\mu$ is a penalty weighting for inconsistency with clause $\alpha$. These indicator functions on phase space encode logical constraints—for instance $\varphi_\alpha(\mathbf{q},\mathbf{p}) = 1$ when the agent's state satisfies clause $\alpha$ of its reasoning system.

\paragraph{Classical form ($H_{\mathrm{reas}}^C$).} Boolean clauses $\varphi_\alpha:T^{*}\mathcal{C}\!\to\!\{0,1\}$, penalty $\mu_\alpha>0$:
\begin{equation}
  H_{\mathrm{reas}}^{C} = \sum_{\alpha=1}^{M} \mu_\alpha\, \delta\bigl(\varphi_\alpha(\mathbf{q},\mathbf{p})-1\bigr).
  \label{eq:H-reas-class-paper3}
\end{equation}
Classical logical propositions typically commute: $\{\varphi_\alpha, \varphi_\beta\}_{PB}=0$ if they depend on distinct configuration variables or are otherwise compatible. The delta function $\delta(\varphi_\alpha(\mathbf{q},\mathbf{p})-1)$ enforces a hard constraint: the energy becomes large unless clause $\alpha$ is satisfied, effectively restricting the system to logically consistent regions of phase space.

\paragraph{Quantum form ($H_{\mathrm{reas}}^Q$)}. In quantum mechanics, logical propositions correspond to projection operators $\Pi_\alpha$ that project onto subspaces where proposition $\alpha$ is 'true'. Unlike classical Boolean functions, these projectors may not commute, leading to fundamental differences in logical inference. We express this as follows via having clauses lift to projectors $\Pi_\alpha$ on $\Hilb_A$:
\begin{equation}
  H_{\mathrm{reas}}^{Q} = \sum_{\alpha=1}^{M} \mu_\alpha\, ( \mathbb{I}-\Pi_\alpha ).
  \label{eq:H-reas-quant-paper3}
\end{equation}
While conventional quantum computation uses unitary sequences $U_a...U_k$, the $( \mathbb{I}-\Pi_\alpha )$ term acts as a penalty to enforce logical constraints during evolution. Contextuality results \cite{kochenspecker1967} may mean a QAGI cannot assign simultaneous, context-independent truth values to all propositions. This may fundamentally alter the nature of logical inference from classical CAGI rule-based systems \cite{goertzel2014}. The Hamiltonian itself is such that its ground states are exactly those classical configurations
(or quantum subspaces) that satisfy \emph{all} logical clauses, because
every added term evaluates to zero there.  
For CAGI the penalties commute, so minimising $H_{\mathrm{reas}}^{C}$ is
order-independent and reproduces ordinary Boolean logic.  
For QAGI, attempting to
simultaneously minimise two incompatible projectors may lead to contextual trade-offs. Contextuality means that the truth value of a proposition can depend on which other propositions are measured first—a phenomenon impossible in classical logic but fundamental to quantum mechanics when dealing with non-commuting observables, potentially requiring strategic choices by QAGI about which logical relationships to evaluate first in complex reasoning chains.
\\
\\
\textbf{Recursion---Self-Reference ($H_{\mathrm{rec}}$)}. Recursive computation and self-reference are fundamental to advanced AI systems, enabling everything from hierarchical reasoning to self-modification capabilities. Recursion can be represented via $H_{\mathrm{rec}}$, models recursive computation and self-reference by tracking the agent's call stack depth—the number of nested function calls or recursive reasoning steps currently active. While actual call stacks are discrete, we approximate stack depth as a continuous coordinate to leverage Hamiltonian mechanics.
\paragraph{Classical form ($H_{\mathrm{rec}}^C$).} The classical form $H^C_{\mathrm{rec}}$ models recursion as a mechanical system with three components: current recursion depth (number of active nested calls) $q_{\mathrm{stk}}$,  $p_{\mathrm{stk}}$ is its conjugate momentum representing the rate of depth change and potential $V_{\mathrm{stk}}(q_{\mathrm{stk}})=\kappa_{\mathrm{s}}q_{\mathrm{stk}}^{2}/2$ capturing the inertia of recursive processes (with $m_s$ a parameter that controls the rate of recursive calls):
\begin{equation}
  H_{\mathrm{rec}}^{C} = \frac{p_{\mathrm{stk}}^{2}}{2m_{\mathrm{s}}} + V_{\mathrm{stk}}(q_{\mathrm{stk}}).
  \label{eq:H-rec-class-paper3}
\end{equation}
Here $q_{\mathrm{stk}}\in\mathbb N$ is the current call-stack depth, $p_{\mathrm{stk}}$ its conjugate momentum, $m_{\mathrm s}$ a parameter that controls how quickly depth can change, and $\kappa_{\mathrm s}$ the spring constant of the harmonic potential $V_{\mathrm{stk}}(q)=\tfrac12\kappa_{\mathrm s}q^{2}$ that energetically penalises deep recursion.  The classical Hamiltonian $H_{\mathrm{rec}}^{C}=p_{\mathrm{stk}}^{2}/2m_{\mathrm s}+V_{\mathrm{stk}}$ therefore aims to keep a CAGI’s stack from growing without bound, while the quantum version $H_{\mathrm{rec}}^{Q}$ comprises clock states $\ket t$, data gates $U_t$, and the halt projector $\Pi_{\!\mathrm{halt}}$. The suitability of this form of Hamiltonian is of course open to debate, but we select it to illustrate the idea that deeper recursion requires more memory and processing resources (represented by higher potential energy), while the momentum term represents the 'inertia' of ongoing recursive computations that resist sudden changes in depth. This models classical sequential processing, where the call stack state is definite. G\"odel machines \cite{schmidhuber2003,steunebrink2011family} involve self-inspection of classical code.

\paragraph{Quantum form ($H_{\mathrm{rec}}^Q$).} The quantum version fundamentally differs by representing the entire computational history in superposition rather than tracking a single definite stack depth. For $\{U_t\}_{t=0}^{L-1}$ on data $\Hilb_d$, clock $\Hilb_c$ (basis $\ket{t},\,t=0,\dots,L$) the quantum form of Hamiltonian is:
\begin{align}
  H_{\mathrm{rec}}^{Q} &= \sum_{t=0}^{L-1} \bigl( \ket{t+1}\!\bra{t}_{\mathrm{clock}}\!\otimes\!U_t + \mathrm{H.c.} \bigr) \nonumber \\
  & \quad + \ket{0}\!\bra{0}_{\mathrm{clock}}\!\otimes (\mathbb{I}-\ket{\psi_0}\!\bra{\psi_0}_{\mathrm{data}}) + \ket{L}\!\bra{L}_{\mathrm{clock}}\!\otimes (\mathbb{I}-\Pi_{\!\mathrm{halt}}).
  \label{eq:H-rec-quant-paper3}
\end{align}
Here, $\ket{t}$ are discrete computational time steps, $U_t$ unitary operations, and H.c. the Hermitian conjugate. The ground state, or history state $\ket{\Psi_{\mathrm{hist}}} \propto \sum_{t=0}^{L} \ket{t} \otimes \bigl(\!\prod_{s=0}^{t-1}\!U_s\bigr)\ket{\psi_0}$, encodes the computation in superposition. Measurement of the current step in the process would collapse this history. This illustrates the difficulties of self-modification and self-inspection for QAGI (albeit these might be addressed with available ancilla). 
\\
\\
\textbf{Learning and Parametric Self-Modification ($H_{\mathrm{learn}}$)}. Learning in AGI systems involves continuous adaptation of internal parameters based on experience. While our earlier induction generator focused on prediction error minimisation, this learning generator models the broader dynamics of parametric self-modification during training and adaptation processes. Let $\boldsymbol{\theta}=(\theta_1,\dots,\theta_d)$ be trainable weights,
$p_{\theta_\ell}$ their conjugate momenta, $m_\ell>0$ effective masses (larger masses correspond to parameters that change slowly, smaller masses allow rapid adaptation)
and $\mathcal L(\boldsymbol{\theta};\mathcal D)$ a differentiable loss
(e.g.\ mean-squared error on the data set $\mathcal D$).  
The learning Hamiltonian (\(\lambda>0\) sets the loss-to-energy scale) is: 
\[
H_{\mathrm{learn}}^{C}(\boldsymbol{\theta},\mathbf p)
=\sum_{\ell=1}^{d}\frac{p_{\theta_\ell}^{2}}{2m_\ell}
\;+\;
\lambda\,\mathcal L(\boldsymbol{\theta};\mathcal D),
\]
\paragraph{Quantum form ($H_{\mathrm{learn}}^Q$).} The quantum formulation requires encoding continuous parameters into discrete qubit states, where each parameter $\theta_\ell$ is represented by the expectation value $\langle Z_\ell \rangle$ of a Pauli-Z operator, with $X_\ell$ and $Z_\ell$ being the standard Pauli matrices for qubit $\ell$. $Z_\ell Z_{\ell'}$
realises Ising couplings \(J_{\ell\ell'}\) that embed the classical
cost landscape. The Ising model, borrowed from statistical physics, uses $Z_\ell Z_{\ell'}$ interactions to encode correlations between parameters, while the transverse fields $g_\ell X_\ell$ create quantum superposition that enables exploration of multiple parameter configurations simultaneously:
\[
H_{\mathrm{learn}}^{Q}
=-\!\sum_{\ell<\ell'}J_{\ell\ell'}Z_\ell Z_{\ell'}
-\!\sum_{\ell}g_\ell X_\ell .
\]
This Hamiltonian embeds the classical loss landscape into quantum spin interactions: the ground state of the Ising terms $-J_{\ell\ell'}Z_\ell Z_{\ell'}$ corresponds to optimal parameter configurations, while the couplings $J_{\ell\ell'}$ encode the loss function's curvature structure. Non-commutation \([X_\ell,Z_\ell Z_{\ell'}]\neq0\) enables tunnelling
through high, narrow barriers, which might accelerate optimisation.
\\
\\
\textbf{Sensing and Environmental Interaction 
($H_{\mathrm{sens}}, H_{\mathrm{env}}$)}.
Sensing the environment can be modelled via $H_{\mathrm{sens}}$ which describes transfers of information from the environment register $\mathsf E$ into an agent sensor register $\mathsf S$.  
In a classical implementation the transfer is a CTC channel that leaves $\mathsf E$ untouched. In the quantum implementation the same coupling entangles a pointer qubit with $\mathsf E$, so reading the pointer realises a QTC channel whose back–action decoheres $\rho_{\mathsf E}$. The pointer qubit $m$ is an ancilla that entangles with environment observable $O_E$ with projective readout decohering $O_E$ off-diagonal terms.
\\
\\
\noindent \textit{Classical form.}
Let $q_{\mathrm{sens}}\in\mathbb R$ be the sensor coordinate inside the agent’s phase space,   
$q_{\mathrm{env}}\in\mathbb R$ the quantity to be read from the environment,   
$P$ the conjugate momentum of $q_{\mathrm{sens}}$,   
and $\kappa>0$ a tunable coupling strength while $\mathbf F(\mathbf q_{\mathrm{env}},\mathbf p_{\mathrm{env}})$ denotes the generalised force $\nabla_{\mathbf q}H_{\mathrm E}^{\mathrm{bare}}$.  
The measurement Hamiltonian is
\begin{equation}
H_{\mathrm{sens}}^{C}=\kappa\,P\,\delta\!\bigl(q_{\mathrm{sens}}-q_{\mathrm{env}}\bigr),
\qquad
H_{\mathrm{env}}^{C}=H_{\mathrm E}^{\mathrm{bare}}-
\mathbf u(t)\!\cdot\!\mathbf F\!\bigl(\mathbf q_{\mathrm{env}},\mathbf p_{\mathrm{env}}\bigr).
\label{eq:H-meas-classical}
\end{equation}
$H_{\mathrm{sens}}^{C}$ vanishes exactly when the sensor value matches
the environmental value, zero energy is expended for a perfect copy and
the Poisson brackets $\{q_{\mathrm{env}},H_{\mathrm{sens}}^{C}\}=0$ show
that $\mathsf E$ is not disturbed.  
The drive term $\mathbf u(t)\cdot\mathbf F$ (with control field
$\mathbf u$ and generalised force $\mathbf F$) keeps the environment
open and classically steerable.
\\
\\
\noindent\textit{Quantum Hamiltonian.}
Write $\ket0_m,\ket1_m$ for the orthogonal pointer states in the
one-qubit sensor register $\Hilb_m$,   
let $O_{\mathsf E}$ be a Hermitian observable on the environment Hilbert
space $\Hilb_E$, and keep the same real constant $\kappa$.  
With $\mathbf A_{\mathsf E}$ a vector of Hermitian operators and $\mathbb I_A$ the identity on the agent’s internal Hilbert space:
\begin{equation}
H_{\mathrm{sens}}^{Q}= \kappa\bigl(\ket1\!\bra0_m\otimes O_{\mathsf E}+ \mathrm{H.c.}\bigr),
\qquad
H_{\mathrm{env}}^{Q}=H_{\mathrm E}^{\mathrm{bare}}-
\mathbf u(t)\!\cdot\!\bigl(\mathbf A_{\mathsf E}\!\otimes\!\mathbb I_A+
\mathrm{H.c.}\bigr).
\label{eq:H-meas-quantum}
\end{equation}
The operator $\ket1\!\bra0_m$ does not commute with its Hermitian
conjugate. This means that  $[H_{\mathrm{sens}}^{Q},H_{\mathrm{sens}}^{Q\,\dagger}]\neq0$ and,
as a consequence, a projective read-out of the pointer implements a
QTC channel whose Lindblad generator
$\mathcal L_{\mathrm{meas}}(\rho)=
-i[H_{\mathrm{sens}}^{Q},\rho]+\gamma\bigl(Z_m\rho Z_m-\rho\bigr)$
suppresses off-diagonal terms of $\rho_{\mathsf E}$ at rate $\gamma\sim\kappa^{2}$. $Z_m$ is the Pauli operator on qubit $m$.
The operator $\ket1 \bra0_m \otimes O_{\mathsf E}$ creates entanglement: when the environment observable $O_{\mathsf E}$ has a particular value, it correlates with flipping the pointer from $\ket{0}_m$ to $\ket{1}_m$, encoding environmental information in pointer-environment correlations. If $[H_{\mathrm{sens}}^{Q},H_{\mathrm{learn}}^{Q}]\neq0$, the same
measurement inevitably perturbs the learning dynamics, and the resulting
agent–environment entanglement can violate Bell inequalities
\cite{Bell1964,Zurek2003}, an effect absent in the commuting classical
model, potentially enabling quantum sensing advantages but also creating fundamental measurement-learning trade-offs impossible in classical AGI.
\\
\\
\textbf{Example Comparison Hamiltonian}\label{sec:example}
To illustrate our approach, we consider the following toy example. Assume the environment is described by a quantum register. A CAGI agent must encode and decode quantum data
via CTQ/QTC interfaces, while a fully quantum QAGI agent that can
also exploit coherent QTQ interactions.  We compose a total Hamiltonian from three subsidiary Hamiltonians $H_{\mathrm{tot}}=H_{\mathrm{sens}}+H_{\mathrm{reas}}+H_{\mathrm{learn}}$.
\\
\\
\textit{QAGI}. The QAGI register consists of a two-qubit policy
$\Hilb_{A_1}\!\otimes\!\Hilb_{A_2}$, a pointer qubit $\Hilb_m$, and the
environment qubit $\Hilb_E$.  Setting \(\hbar=1\),
\begin{align}
H_Q
&=
\kappa\!\!\underbrace{\bigl(\ket1\!\bra0_m\otimes Z_E+\mathrm{H.c.}\bigr)}_{\text{sensing channel}}
+\underbrace{\mu\bigl(\mathbb I-\Pi_{\alpha}\bigr)}_{\text{reasoning error penalty}}
+\underbrace{g X_{A_1}+J Z_{A_1}Z_{A_2}}_{\text{QTQ learning block}},
\label{eq:HQ}
\end{align}
Here $\Pi_{\alpha}=\tfrac12(\mathbb I+Z_m)\otimes\tfrac12(\mathbb I+Z_{A_1})$ enforces $Z_m,Z_{A_1}=+1$. The first term realises a \textit{QTC} measurement: it entangles the
pointer with $Z_E$, and a subsequent pointer read-out transfers the
result to a classical log while decohering $\rho_E$.  
The $\mu$-term operates purely within the quantum formalism: it conditions the system's energy on a projector that fails to commute with the QTC coupling, hence logical consistency is contextual.  
The Ising transverse field block is also QTQ; its non-commutation lets
the policy search landscape be traversed through \emph{tunnelling},
visible as peaks in the quantum Fisher information
$F_\theta(t)=\mathrm{Tr}[L_\theta^2\rho_t]$.
\\
\\
\textit{CAGI}. CAGI possesses only classical registers, so it must encode and
decode quantum data.  Let $q_E\,(=\pm1)$ be the $Z_E$ eigenvalue
obtained by an external QTC reader, $q_m$ the classical sensor bit,
$\theta\in\mathbb R$ a weight, and $p_\theta,m$ as before.  Define the
action bit $q_A=\mathbbm1_{\{\theta>0\}}$.  The Hamiltonian is: \vspace{-0.5em}
\begin{equation}\label{eq:HC}
H_C= \kappa\!\!\underbrace{\delta(q_m-q_E)}_{\text{CTC copy}}
\\[-1pt] 
+\mu\!\!\underbrace{\delta\!\bigl[(1-q_m)q_A-1\bigr]}_{\text{CTC logic}}
\\[-1pt] 
+\frac{p_\theta^{2}}{2m}
+\lambda|\theta|
\\[-1pt] 
+\eta(t)\!\!\underbrace{\bigl[q_A\,Z_E\bigr]}_{\text{CTQ actuator}} 
\end{equation}
The first two deltas are \textit{CTC}—they move only classical bits and
therefore commute with everything else.  
The last line is a time-dependent \textit{CTQ} channel: a classical
action bit $q_A$ is written into the quantum environment
operator $Z_E$ via a control field $\eta(t)$ (e.g.\ a laser pulse that
rotates the obstacle qubit).  There is \emph{no} QTQ term because the
agent cannot maintain coherence; sensing happens by an external QTC
device that prepares $q_E$. The fundamental distinction between CAGI and QAGI lies in commutation: CAGI terms commute completely while QAGI terms do not, creating qualitatively different agent-environment dynamics.  
The CTC copy term \(\kappa\,\delta(q_m-q_E)\) leaves the quantum environment untouched, the Boolean penalties commute so their evaluation order is immaterial, and the weight trajectory \((\theta(t),p_\theta(t))\) follows a smooth deterministic hill–climb.  
By contrast, for QAGI the measurement, reasoning and learning blocks fail to commute.  
A pointer read–out (QTC) entangles then decoheres the obstacle qubit, injecting energy of order \(\kappa\) and shifting the logical penalty because \([H_{\mathrm{sens}}^{Q},H_{\mathrm{reas}}^{Q}]\neq0\); the truth therefore becomes formally context-dependent.\\
\\
\textbf{Conclusion and Discussion}.
 We have proposed a generator-based Hamiltonian framework in which the
total dynamics of an agent are written as a sum of subsidiary
Hamiltonians. For each generator we provided \textit{(i)} a \emph{classical} phase-space
realisation $H_G^C$ acting on a commutative algebra of observables and
\textit{(ii)} a \emph{quantum} operator realisation $H_G^Q$ acting on a
non-commutative von-Neumann algebra. The simplified examples above demonstrate how our framework captures both the computational aspects (via Hamiltonians) and the information-theoretic aspects (via channel types) in a unified description. More complex scenarios may involve coupling and correlations for both CAGI and QAGI. Potential future research avenues include: (i) implementing small-scale agent–in-the-loop experiments on NISQ hardware; (ii) extending many-body and open environments ; and (iii) embedding alignment and safety constraints as additional commuting / non-commuting generators.

%
%
\bibliographystyle{splncs04}
\bibliography{refs-final}

\begin{thebibliography}{10}
\providecommand{\url}[1]{\texttt{#1}}
\providecommand{\urlprefix}{URL }
\providecommand{\doi}[1]{https://doi.org/#1}

\bibitem{aaronson2013quantum}
Aaronson, S.: Quantum computing since Democritus. Cambridge University Press (2013)

\bibitem{Bell1964}
Bell, J.S.: On the {E}instein {P}odolsky {R}osen {P}aradox. Physics Physique Fizika  \textbf{1},  195--200 (1964)

\bibitem{bell_speakable_2004}
Bell, J.: Speakable and {Unspeakable} in {Quantum} {Mechanics}. {Cambridge} {University} {Press}, Cambridge, 2nd edn. (2004)

\bibitem{bennettmaruyama2022b}
Bennett, M.T., Maruyama, Y.: The artificial scientist: Logicist, emergentist, and universalist approaches to artificial general intelligence. In: Artificial General Intelligence. Springer (2022)

\bibitem{chiribella2008quantum}
Chiribella, G., D’Ariano, G.M., Perinotti, P.: Quantum circuit architecture. Physical review letters  \textbf{101}(6),  060401 (2008)

\bibitem{chruscinski_geometric_2004}
Chruściński, D., Jamiołkowski, A.: Geometric phases in classical and quantum mechanics. Springer (2004)

\bibitem{feynman_simulating_1982}
Feynman, R.P.: Simulating physics with computers. International Journal of Theoretical Physics  \textbf{21}(6),  467--488 (Jun 1982)

\bibitem{frankel_geometry_2011}
Frankel, T.: The {Geometry} of {Physics}: {An} {Introduction}. {Cambridge} {University} {Press} (2011)

\bibitem{goertzel2014}
Goertzel, B.: Artificial general intelligence: concept, state of the art, and future prospects. Journal of Artificial General Intelligence  \textbf{5}(1), ~1 (2014)

\bibitem{goertzel2021}
Goertzel, B.: The general theory of general intelligence: A pragmatic patternist perspective (2021)

\bibitem{goertzel2023}
Goertzel, B., Bogdanov, V., Duncan, M., Duong, D., Goertzel, Z., Horlings, J., Ikle', M., Meredith, L.G., Potapov, A., de~Senna, A.L., Suarez, H.S.A., Vandervorst, A., Werko, R.: Opencog hyperon: A framework for agi at the human level and beyond (2023)

\bibitem{goldstein_classical_2002}
Goldstein, H.: Classical {Mechanics}. Pearson Education (Sep 2002)

\bibitem{hall_quantum_2013}
Hall, B.C.: Quantum theory for mathematicians. Springer (2013)

\bibitem{helgason_differential_1979}
Helgason, S.: Differential {Geometry}, {Lie} {Groups}, and {Symmetric} {Spaces}. {ISSN}, Elsevier Science (1979)

\bibitem{hutter2004universal}
Hutter, M.: Universal artificial intelligence: Sequential decisions based on algorithmic probability. Springer Science \& Business Media (2004)

\bibitem{hutter2024introduction}
Hutter, M., Quarel, D., Catt, E.: An Introduction to Universal Artificial Intelligence. CRC Press (2024)

\bibitem{knapp_lie_1996}
Knapp, A.W., Knapp, A.W.: Lie groups beyond an introduction, vol.~140. Springer (1996)

\bibitem{kochenspecker1967}
Kochen, S., Specker, E.P.: The problem of hidden variables in quantum mechanics. Journal of Mathematics and Mechanics  \textbf{17}(1),  59--87 (1967)

\bibitem{manin2007mathematics}
Manin, I.I.: Mathematics as metaphor: Selected essays of Yuri I. Manin, vol.~20. American Mathematical Soc. (2007)

\bibitem{mcmillen2024}
McMillen, P., Levin, M.: Collective intelligence: A unifying concept for integrating biology across scales and substrates. Communications Biology  \textbf{7}(1), ~378 (Mar 2024)

\bibitem{NielsenChuang2010}
Nielsen, M.A., Chuang, I.L.: Quantum Computation and Quantum Information. Cambridge University Press, 10th anniversary edn. (2010)

\bibitem{ozkural2012like}
{\"O}zkural, E.: What is it like to be a brain simulation? In: International Conference on Artificial General Intelligence. pp. 232--241. Springer (2012)

\bibitem{perrier2024quantum}
Perrier, E.: Quantum geometric machine learning. arXiv preprint arXiv:2409.04955  (2024)

\bibitem{potapov2014making}
Potapov, A., Rodionov, S.: Making universal induction efficient by specialization. In: International Conference on Artificial General Intelligence. pp. 133--142. Springer (2014)

\bibitem{preskill_quantum_2021}
Preskill, J.: Quantum computing 40 years later. arXiv:2106.10522  (2021)

\bibitem{RussellNorvig2020}
Russell, S., Norvig, P.: Artificial Intelligence: A Modern Approach. Pearson, 4th edn. (2020)

\bibitem{sakurai2020modern}
Sakurai, J.J., Napolitano, J.: Modern quantum mechanics. Cambridge University Press (2020)

\bibitem{schmidhuber2003}
Schmidhuber, J.: G{\"o}del machines: Self-referential optimal universal problem solvers. arXiv preprint cs/0309048  (2003)

\bibitem{solomonoff1964}
Solomonoff, R.J.: A formal theory of inductive inference. part i \& ii. Information and Control  \textbf{7}(1--2),  1--22, 224--254 (1964)

\bibitem{sole2019}
Solé, R., Moses, M., Forrest, S.: Liquid brains, solid brains. Philosophical Transactions of the Royal Society B: Biological Sciences  \textbf{374}(1774),  20190040 (2019)

\bibitem{steunebrink2011family}
Steunebrink, B.R., Schmidhuber, J.: A family of g{\"o}del machine implementations. In: Artificial General Intelligence: 4th International Conference, AGI 2011, Mountain View, CA, USA, August 3-6, 2011. Proceedings 4. pp. 275--280. Springer (2011)

\bibitem{sunehag2012optimistic}
Sunehag, P., Hutter, M.: Optimistic aixi. In: Artificial General Intelligence: 5th International Conference, AGI 2012, Oxford, UK, December 8-11, 2012. Proceedings 5. pp. 312--321. Springer (2012)

\bibitem{veness2012ensemble}
Veness, J., Sunehag, P., Hutter, M.: On ensemble techniques for aixi approximation. In: International Conference on Artificial General Intelligence. pp. 341--351. Springer (2012)

\bibitem{watrous_theory_2018}
Watrous, J.: The {Theory} of {Quantum} {Information}. {Cambridge} {University} {Press} (2018)

\bibitem{wootters_single_1982}
Wootters, W.K., Zurek, W.H.: A single quantum cannot be cloned. Nature  \textbf{299},  802--803 (Oct 1982)

\bibitem{Zurek2003}
Zurek, W.H.: Decoherence, einselection, and the quantum origins of the classical. Reviews of Modern Physics  \textbf{75}(3),  715--775 (2003)

\end{thebibliography}

\newpage
\appendix
\section*{Technical Appendices}
\section{Diagrams}


\begin{figure}[ht!]
\centering
\footnotesize               

\resizebox{0.5\textwidth}{!}{
\begin{circuitikz}[every node/.style={font=\footnotesize}]
\draw[rounded corners,fill=orange!20] (3.25,7.5) rectangle node {\scriptsize CAGI} (5,6.5);
\draw[rounded corners,fill=gray!15 ] (6.25,8.75) rectangle node {\scriptsize $E_C$} (8,7.75);
\draw[rounded corners,fill=gray!15 ] (6.25,6.25) rectangle node {\scriptsize $E_Q$} (8,5.25);
\draw (4,7.5)--(4,8.5);
\draw[->] (4,8.5)--(6.25,8.5) node[pos=.5,fill=white]{\scriptsize CTC};
\draw (6.25,8)--(4.25,8)      node[pos=.5,fill=white]{\scriptsize CTC};
\draw[->] (4.25,8)--(4.25,7.5);
\draw (4,6.5)--(4,5.5);
\draw[->] (4,5.5)--(6.25,5.5) node[pos=.5,fill=white]{\scriptsize CTQ};
\draw (6.25,6)--(4.25,6)      node[pos=.5,fill=white]{\scriptsize QTC};
\draw[->] (4.25,6)--(4.25,6.5);
\draw[->] (6.75,7.75)--(6.75,6.25) node[pos=.5,fill=white]{\scriptsize CTQ};
\draw[->] (7.5,6.25)--(7.5,7.75)   node[pos=.5,fill=white]{\scriptsize QTC};
\end{circuitikz}}
\caption{\scriptsize Classical agent (CAGI) interacting via CTC, CTQ or QTC maps with classical $E_C$ or quantum $E_Q$ environments} \label{fig:CAGI}

\vspace{1.2em}  

\resizebox{0.5\textwidth}{!}{%
\begin{circuitikz}[every node/.style={font=\scriptsize}]
\draw[rounded corners,fill=blue!15 ] (3.25,7.5) rectangle node {\scriptsize QAGI} (5,6.5);
\draw[rounded corners,fill=gray!15 ] (6.25,8.75) rectangle node {\scriptsize $E_C$} (8,7.75);
\draw[rounded corners,fill=gray!15 ] (6.25,6.25) rectangle node {\scriptsize $E_Q$} (8,5.25);
\draw (4,7.5)--(4,8.5);
\draw[->] (4,8.5)--(6.25,8.5) node[pos=.55,fill=white]{\scriptsize QTC};
\draw (6.25,8)--(4.25,8)      node[pos=.55,fill=white]{\scriptsize CTQ};
\draw[->] (4.25,8)--(4.25,7.5);
\draw (4,6.5)--(4,5.5);
\draw[->] (4,5.5)--(6.25,5.5) node[pos=.55,fill=white]{\scriptsize QTQ};
\draw (6.25,6)--(4.25,6)      node[pos=.55,fill=white]{\scriptsize QTQ};
\draw[->] (4.25,6)--(4.25,6.5);
\draw[->] (6.75,7.75)--(6.75,6.25) node[pos=.5,fill=white]{\scriptsize CTQ};
\draw[->] (7.5,6.25)--(7.5,7.75)   node[pos=.5,fill=white]{\scriptsize QTC};
\end{circuitikz}}
\caption{\scriptsize Quantum agent (QAGI) interacting via QTC, CTQ or QTQ maps.} \label{fig:QAGI}
\vspace{-1em}
\end{figure}

\section{Hamiltonian Dynamics}
\textbf{Classical Evolution} Evolution is described by Hamilton's equations:
\begin{equation}
    \dot{q}_i = \frac{\partial H_C}{\partial p_i}, \quad \dot{p}_i = -\frac{\partial H_C}{\partial q_i}.
\end{equation}
This can be expressed more abstractly using the Poisson bracket. For two observables $f, g$, their Poisson bracket is:
\begin{equation}
    \{f, g\}_{PB} = \sum_{i=1}^n \left( \frac{\partial f}{\partial q_i}\frac{\partial g}{\partial p_i} - \frac{\partial f}{\partial p_i}\frac{\partial g}{\partial q_i} \right).
\end{equation}
The time evolution of any observable $f$ is then given by $\dot{f} = \{f, H_C\}_{PB}$. \noindent
When \(\{f,g\}_{\mathrm{PB}} = 0\) the Poisson bracket vanishes, so the observables \(f\) and \(g\) \emph{commute} and their values can, in principle, be fixed simultaneously with arbitrary accuracy. Classical logic and computation implicitly assume this independence: the truth of one proposition (or the content of one register) leaves another untouched unless an explicit coupling Hamiltonian \(H_C\) is present. Consequently, for a classical AGI we write the control Hamiltonian as a direct sum
\[
  H_C \;=\; \sum_{k} H_{C,k},
\]
where each term \(H_{C,k}\) drives a distinct functional block—learning (e.g.\ gradient-descent updates~\cite{sunehag2012optimistic}), reasoning (e.g.\ a Hopfield-network energy or constraint-satisfaction term), or sensorimotor exchange. The mutual commutativity of these blocks, and of the variables they address, underpins the semantics of classical computation.

%
In information theoretic terms, classical mechanical dynamics of CAGI can be expressed as follows. Let $\{\mathcal R_i\}_{i=1}^{n}$ be the classical registers of the agent,
each described by a \emph{commutative} von~Neumann algebra
$\mathcal V_i=L^{\infty}(\Omega_i,\mu_i)$.
A micro-state of the whole agent–environment system is therefore a point
$(\mathbf q,\mathbf p)\in\mathcal M=T^{*}\!\mathcal C$ with
\[
q_i := X_i(\omega_i), \qquad
p_i := M_i\,\dot X_i(\omega_i),
\]
where $X_i\in\mathcal V_i$ is the random variable realised by register
$\mathcal R_i$ and $M_i$ is an information-theoretic weight term (e.g.\ an inverse learning rate or buffer capacity). The classical Hamiltonian functional
$H_C\colon\mathcal M\to\mathbb R$
involves coordinates:
\begin{equation}\label{eq:class-ham-eq}
  \dot q_i \;=\;\pdv{H_C}{p_i}, 
  \qquad
  \dot p_i \;=\;-\pdv{H_C}{q_i},
\end{equation}
but now \eqref{eq:class-ham-eq} is understood to act on \emph{probability
densities} $f_t(\mathbf q,\mathbf p)$ pushed forward by the
CTC channel
$\mathrm{CTC}_t\!:L^{\infty}(\Omega)\!\to\!L^{\infty}(\Omega)$. For any pair of observables
$f,g\in\bigoplus_i\mathcal V_i$
we retain the Poisson bracket:
\[
  \{f,g\}_{\mathrm{PB}}
  \;=\;
  \sum_{i=1}^{n}
    \Bigl(\pdv{f}{q_i}\pdv{g}{p_i}
          -\pdv{f}{p_i}\pdv{g}{q_i}\Bigr),
\]
so the time derivative of $f$ is
$\dot f=\{f,H_C\}_{\mathrm{PB}}$.
In information terms
$\{f,g\}_{\mathrm{PB}}=0$
\emph{iff} the corresponding
\emph{classical channels} commute.


\subsection{Quantum Hamiltonian dynamics}
Upon shifting to a \emph{quantum} substrate, the AGI’s state lives as a vector \(\ket{\psi}\) in a Hilbert space \(\Hilb\) (or, more generally, as a density operator \(\rho\) on \(\Hilb\)).  Observables correspond to self-adjoint operators \(A\) acting on that space, and evolution follows the Schrödinger–von Neumann equation
\[
  i\hbar\,\frac{\dd}{\dd t}\rho(t)=\,[H_Q,\rho(t)],
\]
where \([A,B]\equiv AB-BA\) is the \emph{commutator}—the quantum analogue of the Poisson bracket—and \(H_Q\) is the total quantum Hamiltonian.  The critical algebraic shift from the classical picture is that operators need not commute: when \([A,B]\neq0\), simultaneous precise values are forbidden, giving rise to distinctively quantum effects discussed below.  For a quantum AGI we likewise decompose
\[
  H_Q \;=\; \sum_{k} H_{Q,k},
\]
each \(H_{Q,k}\) generating a functional capability—learning, reasoning, perception, actuation, and so forth.  Now, however, the \emph{commutation relations} among these generators, and with other key observables, govern behaviour: if the learning term \(H_{Q,\mathrm{learn}}\) fails to commute with the sensing term \(H_{Q,\mathrm{sens}}\), then observation can disturb learning (and vice versa) in a way with no classical counterpart.  Such non-commutative structure underlies quantum phenomena like entanglement and contextuality and therefore reshapes the semantics of computation in a quantum-enabled AGI.
This non-commutativity is fundamental and has profound consequences which may be a constraint or benefit.

For a quantum AGI we again write the control Hamiltonian as a sum of functional generators,
\[
  H_Q \;=\; \sum_k H_{Q,k}.
\]
Each \(H_{Q,k}\) is now an \emph{operator}, so the commutators among these terms—and with other observables—govern the agent’s evolution.  If, say, the learning generator \(H_{Q,\mathrm{learn}}\) fails to commute with the sensing generator \(H_{Q,\mathrm{sens}}\), environmental measurement can disturb learning (and vice versa) in a manner with no classical analogue.  This non-commutative architecture underlies quantum hallmarks such as entanglement and contextuality, which may represent either valuable resources or formidable challenges for a QAGI.

\subsection{Quantum information formulation} 
In quantum information terms, transitioning to quantum involves replacing every classical register $\mathcal V_i=L^{\infty}(\Omega_i)$ by a
\emph{non-commutative} von Neumann algebra
$\mathcal V_i=B(\mathcal H_i)$ acting on a Hilbert space
$\mathcal H_i$. The full agent–environment is reflected by the tensor
algebra
\(
\mathcal V=\bigotimes_i\mathcal V_i\subseteq\Bcal(\Hilb),
\)
with $\mathcal H=\bigotimes_i\mathcal{H}_i$. States are represented by density operators
\(
\rho\in\mathcal D(\Hilb)
=\{\rho\ge0,\ \Tr\rho=1\},
\)
and a observable is an element
$A\in\mathcal V$. When the evolution is closed and reversible the channel on
$\mathcal V$ is the adjoint action of a unitary $U_t$:
\[
  \Phi_t^{(\text u)}(A)=U_t^{\dagger}AU_t,
  \qquad
  U_t=\exp\!\bigl(-\tfrac{i}{\hbar}H_Q t\bigr),
\]
where the \emph{Hamiltonian operator}
$H_Q\in\mathcal V$ is the quantum analogue
of~$H_C$. In Schr\"odinger form this yields the familiar
\begin{align}
  i\hbar\dot{\rho}(t)
    =\bigl[H_Q,\rho(t)\bigr],
  \label{eq:qschrod}
\end{align}
which is the generator
$\mathcal L_{H_Q}=-\tfrac{i}{\hbar}[H_Q,\cdot]$
of a one-parameter group of
QTC channels. Realistic AGI modules interact with—and are monitored by—their environment, so the fundamental dynamical object is a
quantum channel
\(
\Phi_t=\exp(t\mathcal L),
\)
with Lindblad superoperator:
\begin{align}
  \mathcal L(\rho)
  =-\frac{i}{\hbar}[H_Q,\rho]
   +\!\sum_\alpha
      \Bigl(L_\alpha\rho L_\alpha^\dagger
            -\tfrac12\{L_\alpha^\dagger L_\alpha,\rho\}\Bigr) \label{eqn:lindblad}
\end{align}
where the $L_\alpha$’s represent QTC measurement-and-feedback registers).

\end{document}